\documentclass[lettersize,journal]{IEEEtran}
\usepackage{amsmath,amsfonts}
\usepackage{algorithmic}
\usepackage{algorithm}
\usepackage{array}
\usepackage[caption=false,font=normalsize,labelfont=sf,textfont=sf]{subfig}
\usepackage{textcomp}
\usepackage{stfloats}
\usepackage{url}
\usepackage{verbatim}
\usepackage{graphicx}
\usepackage{cite}
\usepackage{float}
\usepackage{booktabs}
\usepackage{hyperref}
\usepackage{xcolor}

\newtheorem{definition}{Definition}[section]
\newcommand{\D}{\displaystyle}
\newcommand{\T}{^\top}
\begin{document}

\title{A New Mathematical Optimization-Based Method for the $m$-invariance Problem}

\author{Adrian Tobar, Jordi Castro, and Claudio Gentile
\thanks{Jordi Castro and Adrian Tobar are with the Universitat Polit\`ecnica de
    Catalunya, Barcelona, Catalonia. E-mail:\{jordi.castro, adrian.tobar\}@upc.edu.}
\thanks{Claudio Gentile is with the Consiglio Nazionale delle Ricerche, Istituto di Analisi dei Sistemi ed Informatica  "A. Ruberti" (IASI-CNR), Roma,
  Italia. E-mail:gentile@iasi.cnr.it.}
\thanks{
Author Adrian Tobar had the original idea of updating the method in
\cite{CasGenSpa22a} 
to the m-invariance and $\tau$-safety problem. All the authors equally contributed in the implementation of the
method and in writing the paper.
}
\thanks{\copyright 2023 IEEE. Personal use of this material is permitted. Permission
from IEEE must be obtained for all other uses, in any current or future
media, including reprinting/republishing this material for advertising or
promotional purposes, creating new collective works, for resale or
redistribution to servers or lists, or reuse of any copyrighted
component of this work in other works
}
}


\markboth{}{}

\IEEEpubid{}

\maketitle

\begin{abstract}
The issue of ensuring privacy for users who share their personal information has been a growing priority in a business and scientific environment where the use of different types of data and the laws that protect it have increased in tandem. Different technologies have been widely developed for static publications, i.e., where the information is published only once, such as $k$-anonymity and $\epsilon$-differential privacy. In the case where microdata information is published dynamically, although established notions such as $m$-invariance and $\tau$-safety already exist, developments for improving utility remain superficial. 

We propose a new heuristic approach for the NP-hard combinatorial problem of
$m$-invariance and $\tau$-safety, which is based on a mathematical optimization
column generation scheme. The quality of a solution to $m$-invariance and
$\tau$-safety can be measured by the Information Loss (\emph{IL}), a value in
[0,100], the closer to 0 the better. We show that our approach
improves by far current heuristics, providing in some instances
solutions with \emph{IL}s of 1.87, 8.5 and 1.93, while the state-of-the art methods
reported \emph{IL}s of 39.03, 51.84 and 57.97, respectively.
\end{abstract}


\section{Introduction}
\IEEEPARstart{T}{he} statistical disclosure control \cite{SDC_book} field is devoted to the private-preserving publication of multiple forms of data. In the microdata publication, a table with information at the individuals level is published. The existing mechanisms for protecting privacy and anonymity of respondents, that is, the users that shared their data, can be broadly classified by two main properties: the data publishing scenario and how they achieve privacy.

Syntactic notions are those that enforce a particular structure on the dataset. On the other hand, semantic notions are those that base their privacy on enforcing certain properties on the algorithms anonymizing the data. Since the inception of statistical disclosure control the most studied publishing scenario was the static data release. Examples of syntactic notions for this framework are $k$-anonymity \cite{Kanon-Massachussets,kanon_original},  l-diversity \cite{Machanavajjhala06ICDE} and t-closeness \cite{Li07ICDE,closeness_Forne} and examples of semantic notions are $\epsilon$-differential privacy \cite{Dwork06A,Dwork10STOC,Dwork14A} and their variations. A notable difference between each approach is that syntactic methods assume a classification of microdata in two types, quasi-identifiers, that is, data that is not sensitive to the users but may be used to partially identify them (age, sex, weight, marital status,...) and sensitive data, i.e., the information that, if associated with a user, would  violate their privacy (medical records, criminal history, salary,...). Broadly speaking semantic methods assume stronger attackers and achieve better privacy guarantees but at the expense of worsening significantly data utility. On the other hand syntactic methods achieve a better trade-off between privacy and utility at the cost of assuming rigid attacker with limited information.

With the interest of using alternative data structures a new umbrella of publishing scenarios has appeared, in particular dynamic scenarios. These scenarios are defined by allowing editions of data and the partial or total republication of data in several independent publications. 

Continuous data publishing \cite{2006} is a dynamic framework where a dataset is periodically published and in-between publications it is updated via insertion of new tuples, deletion of existing ones, updates of microdata and reinsertion of previously deleted tuples. There are three levels of dynamicity: incremental, the dataset can be increased adding new users, i.e., new rows; external dynamic, rows can be added and deleted but a row from a deleted user cannot be reinserted; fully dynamic, additions, deletions, reinsertions and updates of microdata are possible.

Since the first proposal for continuous data publishing  due to Byun
et. al. \cite{2006} several notions and algorithms have been proposed
\cite{2007,2008-0,2009-x,2013-1,2019-1,2022_cach,2022_fuzzy} to handle various
attackers and publishing scenarios. Among them $m$-invariance \cite{2007}
appeared as the first clear notion that bounds the capacity of an
attacker. However $m$-invariance was limited to dynamic datasets, i.e.,
datasets which only update inserting and deleting tuples. To overcome these
limitations,  $\tau$-safety \cite{2013-1,2017} was proposed. Fundamentally
$\tau$-safety strengthens $m$-invariance at the expense of stronger
assumptions. The more recent advancements in continuous data publishing
\cite{2019-1,mDistincImprovement2019,2022_cach} are slight improvements of $\tau$-safety and
their implementations with the exception of \cite{2022_fuzzy} which present a
new enforcement algorithm based on fuzzy clustering. Nevertheless the study of
enforcing $m$-invariance and $\tau$-safety  has been barely non-existent and
in most cases only improvements of the original algorithm of
$m$-invariance have been carried out. Furthermore, no deep analysis of the combinatorial problem of obtaining $m$-invariance has been performed so far.

$m$-Invariance and $\tau$-safety are related to the microaggregation problem
\cite{DomingoF}, a privacy preserving technique that guarantees
\emph{$k$-anonymity}. Briefly, given a set of points, the goal of
microaggregation is to partition them into clusters of a minimum size $k$ ($k$
being a parameter of the problem) that minimize the \emph{information loss
(\emph{IL})} (to be defined later in Section \ref{sec:IPmodels}). A partition
satisfying the constraint on cluster cardinality is referred as
\emph{feasible clustering}, and, of all the feasible clusterings, the one
minimizing \emph{IL} is named the \emph{optimal clustering}. Microaggregation
is known to be a NP-hard problem \cite{NP-Hard}.

The purpose of $m$-invariance is also to find an optimal clustering ($m$ being
the minimum cluster cardinality) with the additional constraint that two points 
in the same cluster can not have the same value for a particular (sensitive)
attribute. For instance, if this particular attribute is named the ``color''
of the point, $m$-invariance finds an optimal clustering where all the points
of a cluster have a different color. If all the points of the dataset have
initially a different color, then $m$-invariance reduces to microaggregation,
and it is thus also a NP-hard problem.

\section{$m$-Invariance and $\tau$-safety}

$m$-Invariance and $\tau$-safety are privacy notions designed to upper bound the probability that an attacker can correctly link a sensitive attribute to a user participating in a dynamic dataset. To present them we first introduce the necessary definitions.

A \emph{dataset} $T$ is a $p\times t$ matrix whose element $i,j$ provides the value of
the attribute $V_j$ of user $i$. The attributes $V_j$ with $1\leq j< t$ are
quasi-identifiers and $V_t$ is considered to be the sensitive attribute. 

The classes of a dataset $T$ are each set of the partition of the rows of $T$ in disjoint subsets such that all rows on each class have common quasi-identifiers, particularly, a class $Q$ is a non-empty subset of users with common quantifiers. We denote as $SD(Q)$ the signature of $Q$, that is, the set of sensitive attributes of a class $Q$.

In general we denote $T$ to refer to a dataset and $T^*$ to refer to its anonymized version. If multiple publications of a changing dataset $T$ are done, we denote as $\textbf{T}=\{T_1,...,T_n\}$ the set of versions of $T$ before each publication and $\textbf{T}^*=\{T_1^*,...,T_n^*\}$ to the set of publications.

The row of microdata $t$, from now on tuple, of a user can belong to several versions of $\mathbf{T}$ and $\mathbf{T^*}$. We denote as \emph{lifespan} of a tuple $h$ as a set $[x,y]=\{x,x+1,...,y\}$ that satisfies $h\in T^*_{i}$ for all $i\in [x,y]$ and $h\notin T_{x-1},T_{y+1}$. If a tuple is deleted and reinserted later, then it can have more than one lifespan. We also define $Q(h,T^*)$ as the class of $h$ in $T^*$.

We say that a dataset has arbitrary updates if whenever a change of microdata is performed, it was not dependant on the previous values. 

With this previous definitions we are now able to state the definitions of $m$-invariance.

\begin{definition}
($m$-invariance) A dataset $T^*$ is \emph{m-unique} if each class in $T$ contains at least $m$ tuples, and all tuples in the class have different sensitive attributes. Let $\textbf{T}^*=\{T_1^*,...,T_n^*\}$ be the distinct publications of an external dynamic dataset, then $\textbf{T}^*$ is \emph{m-invariant} if the following conditions hold:
\begin{itemize}
\item $T^*_i$ is $m$-unique for all $i\in [1,n]$.
\item For any tuple $h$ with lifespan $[x,y]$ it is satisfied $SD(Q(h,T_i^*))=SD(Q(h,T_j^*))$ for all $i,j\in [x,y]$.
\end{itemize}
\end{definition}


We state now the definition of $\tau$-safety.

\begin{definition}
Let $\textbf{T}^*=\{T_1^*,...,T_n^*\}$ be the distinct publications of a fully dynamic dataset with arbitrary updates, then $\textbf{T}^*$ is \emph{$\tau$-safe} if the following conditions hold:  
\begin{itemize}
    \item $\textbf{T}^*$ is $m$-invariant.
    \item For any tuple $h$ with lifespans $[x,y],[z,t]$ it is satisfied $SD(Q(h,T_y^*))=SD(Q(h,T_z^*))$.
\end{itemize}
\end{definition}

The motivation behind these definitions is ensuring that the republication of
data cannot allow the attacker to deduce sensitive information of any user
participating in the dataset.

We illustrate previous ideas with the following example of intersection attack.
Assume an attacker is searching information of a participant with
$age=18$. From Table~\ref{fig:first}  deduces that it has sensitive value HIV
or FLU and from the Table~\ref{fig:second_diverse} that it has HIV or
ACNE. Intersecting both cases, the attacker deduces that the attacked tuple
has HIV. Such attacks are avoidable using $m$-invariance, in this case,
publishing Table~\ref{fig:second_invariant} instead of
Table~\ref{fig:second_diverse}. 

\begin{table}[ht]
\centering
\caption{First 2-diverse publication.}
\label{fig:first}
\begin{tabular}{ccc}\\ \hline
Id & AGE & S.V. \\ \hline
1  & [18-20]   & HIV  \\
2  &  [18-20]   & FLU \\ \hline 
\end{tabular}
\bigskip
\caption{Second 2-diverse but not 2-invariant publication.}
\label{fig:second_diverse}
\begin{tabular}{ccc}\\ \hline
Id & AGE & S.D. \\ \hline
1  & [18-19]   & HIV  \\
3  & [18-19]   & ACNE \\ 
2 & [20-21] & FLU \\
4 & [20-21] & COUCH \\ \hline 
\end{tabular}
\bigskip
\caption{Second 2-invariant publication.}
\label{fig:second_invariant}
\begin{tabular}{ccc}\\ \hline
Id & AGE & S.D. \\ \hline
1  & [18-20]   & HIV  \\
2  & [18-20]   & FLU \\ 
3 & [19-21] & ACNE \\
4 & [19-21] & COUCH \\ \hline 
\end{tabular}
\end{table}

\subsection{Enforcing $m$-invariance and $\tau$-safety}

Most proposals to enforce $m$-invariance and $\tau$-safety use the same bucketization algorithm: classification, balancing, assignment and partitioning. We present now the main ideas behind these algorithms and where does our proposal improve the state of the art.

A bucket is a data structure which uses the key values as the indices of the buckets, for instance, given a bucket $B$ then $B[sd]$ are the tuples in $B$ with sensitive value $sd$. A bucket $B$ has signature $SD(B)$, the set of its keys. A bucket is balanced if for all keys it has the same number of tuples, otherwise it is unbalanced. The bucket algorithms to obtain $m$-invariance or $\tau$-safety proceeds as follows.

\begin{itemize}
    \item Classification: the tuples in the dataset are categorized as new
      (never published), and old (previously published). This yields two datasets $T_{new}$ and $T_{old}$. Data from $T_{old}$ is stored in multiple bucket datasets in the following manner: for each tuple $h$, if a bucket $B$ with signature $SD(Q(h,T^*))$\footnote{Signature of the class of the last publication of $h$.} exists, add $h$ to $B$, otherwise create a bucket with that signature and add $h$ to it.
    \item Balancing: for each bucket created in the classification step, if it is unbalanced, add tuples from $T_{new}$ until it is balanced, if none available add counterfeits.
    \item Assignment: divide the remaining tuples of $T_{new}$ in balanced buckets of at least signature size $m$.
    \item Partitioning: for each bucket $B$ divide it in groups of tuples appropriately. Generalize each group into a class and publish the dataset.
\end{itemize}
This algorithm structure allows for a republication which does not increase drastically in time complexity as new versions are published.

The bulk of the utility lost is due to steps of assignment and
partitioning. Only one method exists in the literature for the assignment
phase presented by Xiao and Tao in \cite{2007}. Partitioning has two versions,
being  \cite{2013-1,2017} the only improvements of the original version in
\cite{2007}. Our work is an alternative method that jointly performs the
assignment and partitioning steps in a single stage. Performing assignment and
partitioning together drastically improves the quality (i.e., reduction of
\emph{IL}) of the solutions computed.

\section{Integer optimization  models}
\label{sec:IPmodels}

Here we adapt to $m$-invariance a 
formulation inspired by the clique partitioning problem with minimum clique
size of \cite{JiMit07}. Defining as ${\cal C}^*= \{ C \subseteq
\{1,\dots,p\}: m \le |C| \le 2m-1, c(i)\neq c(j) \mbox{ for } i,j \in C \}$ the set of feasible clusters, where $c(i)$ is the color of element $i$, 
the $m$-invariance problem can be formulated as:
\begin{equation}
\label{eq:CGprob}
  \begin{array}{rll}
 \min \quad & \D \sum_{C \in \mathcal{C}^*} w_{C}x_{C} \\
 \hbox{s. to} \quad &\D\sum_{C \in \mathcal{C}^*: i \in C} x_C = 1  \quad &  i\in\{1,\dots,p\} \\
 & x_{C} \in \{0,1\} & {C} \in \mathcal{C}^*,
  \end{array}
\end{equation}
where $x_C=1$ means that feasible cluster $C$ appears in the
$m$-invariance provided solution, and the constraints guarantee that all
the points are covered by some feasible cluster, and only once, that is, a
point can not belong to two different clusters, thus having a partition of
$\{1,\dots,p\}$.

A widely used measure for evaluating the quality of a clustering is its
spread or {\em sum of squared errors} ($SSE$) \cite{DomingoF}:
\begin{equation}
SSE = \sum_{C\in{\cal C}^*:x_C=1} SSE_C
\label{eq:SSE_definition}
\end{equation}
where $SSE_C$ is the spread of cluster $C$ which is defined as
\begin{equation}
SSE_C = \sum_{i\in C}(a_i - \overline{a}_C)\T (a_i - \overline{a}_C),
\label{eq:SSE_C}
\end{equation}
$a_i$ being a point of the cluster and $\bar{a}_C=\frac{1}{|C|} \sum_{i \in
C} a_i$ its centroid.

The cost $w_C$ of cluster $C$ in the objective function of \eqref{eq:CGprob} is
\begin{equation}
  w_C = \frac{1}{2|C|} \sum_{i \in C} \sum_{j \in C} D_{ij},
  \label{eq:clusterweight}
\end{equation}
where $D_{ij} = (a_i - a_j)\T (a_i - a_j)$. Using, for every cluster $C$,
the following well-known equivalence (see \cite{CasGenSpa22a}):
\begin{equation}
  \sum_{i \in C} (a_i - \bar{a}_C)\T (a_i - \bar{a}_C) =
  \frac{1}{2|C|} \sum_{i \in C} \sum_{j \in C}  (a_i - a_j)\T (a_i - a_j),
\end{equation}
we have that $w_C=SSE_C$, and then the objective function of \eqref{eq:CGprob}
equals $SSE$.

Information loss \emph{IL} is an equivalent measure to $SSE$, defined as
\begin{equation}
IL= \frac{SSE}{SST}\cdot 100,
\end{equation}
where $SST$ is the total sum of squared errors for all the points:
\begin{equation}
SST= \sum_{i=1}^p (a_i - \bar a)\T (a_i - \bar a)\quad \hbox{ where } \bar a= \frac{\sum_{i=1}^p a_i}{p}.
\end{equation}
\emph{IL} always takes values within the range $[0,100]$; the smaller the \emph{IL}, the better
the clustering. Therefore, the optimal solution of \eqref{eq:CGprob} provides
the feasible clustering that minimizes the information loss.

It is worth noting that in the definition of ${\cal C}^*$
only clusters of cardinality $|C| \le 2m-1$ are considered, since, as proved in 
\cite{DomingoF}, a cluster of cardinality $|C|= 2m$ can be divided in two
smaller clusters of $m$ points, thus improving the \emph{IL}.

The number of feasible clusters in ${\cal C}^*$---that is, the number of variables in the
optimization problem \eqref{eq:CGprob}--- can be huge, so
its direct solution by optimization methods is unpractical
at least for large sizes.
Therefore we resort to heuristics based on two ingredients:
\begin{itemize}
\item decomposition,
\item column generation,
\end{itemize}
that will be detailed in next two sections

\section{Decomposition}
\label{sec:decomposition}
The decomposition heuristic is an extension of the heuristic initially developed for
the microaggregation problem in \cite{CasGenSpa22b}.

The decomposition heuristic is based on partitioning the initial set of points
${\cal P}= \{1,\dots,p\}$ in $s$ subsets ${\cal P}_k, k=1,\dots,s$,
such that $\cup_{k=1}^s {\cal P}_k= {\cal P}$, and ${\cal P}_k \cap {\cal P}_l= \emptyset$
for all $k,l:k\not=l$. This initial partitioning is obtained by first finding a feasible clustering using the two existing heuristics for $m$-invariance (namely, the classical approach \cite{2007} and the $\tau$-safety proposal \cite{2013-1,2017}). Then, points in different clusters of this initial
clustering are sequentially added, obtaining the initial partitioning ${\cal
  P}_k$, $k=1,\dots,s$.

For each subset ${\cal P}_k$, $k=1,\dots,s$, we then consider the
(smaller) optimization problem \eqref{eq:CGprob} replacing the feasible set
${\cal C}^*$ by ${\cal C}^*_k=\{ C \subseteq{\cal P}_k: m \le |C| \le 2m-1, c(i)\neq c(j)
\mbox{ for } i,j \in C \}$. These $s$ optimization problems, though smaller
than the original problem \eqref{eq:CGprob}, may still have a very large number of
optimization variables and are solved using the column generation technique
described below in Section \ref{sec:CG}. Each of the $s$ optimization problems
will provide a set of feasible clusters ${\cal O}_k \subseteq {\cal C}^*_k$
for the subset of points ${\cal P}_k$, and therefore its union ${\cal O}=
\cup_{k=1}^s {\cal O}_k$ will be a feasible clustering for ${\cal P}$
(suboptimal, but in general of good quality---that is, small information loss).

Additionally, the feasible clustering ${\cal O}$ is further improved by
applying a local search heuristic based on a two-swapping procedure. In short,
this procedure analyzes all the feasible swappings between two points $i$
and $j$ located in different clusters $C_i$ and $C_j$, such that  $c(i)\neq c(h)$
for each $h \in C_j$ and $c(j) \neq c(h)$ for each $h \in C_i$, performing
the swapping of the pair $(i,j)$ that minimizes the objective function of 
\eqref{eq:CGprob}. 
This is repeated until there is no improvement in the objective function.
The cost per iteration of this procedure is $O(p^2/2)$.

The two-swapping heuristics can also be optionally used to obtain the initial
partitioning ${\cal P}_k$, $k=1,\dots,s$, of points. Indeed, two-swapping can be applied to the initial
clustering found by the $m$-invariance heuristics, obtaining a new clustering with a smaller
objective function. This new clustering is then used to obtain the initial partitioning.

The main steps of the decomposition heuristic can be summarized as follows:

\begin{enumerate}
\item Apply $m$-invariance heuristics to get an initial feasible clustering.
\item Optionally, apply the two-swapping heuristic to this initial feasible
  clustering.
\item Compute the initial partitioning ${\cal P}_k$, $k=1,\dots,s$, of points
  from the initial feasible clustering.
\item Apply the column generation optimization algorithm to each set of points 
 ${\cal P}_k$, obtaining a feasible clustering ${\cal O}_k$ 
 for all $k=1,\dots,s$. Note that this step can be performed in parallel for
 all the sets $k=1,\dots,s$.
 \item Compute ${\cal O}=\cup_{k=1}^s {\cal O}_k$, which is a feasible
   clustering for ${\cal P}$.
 \item Finally, apply the two-swapping heuristic to the feasible clustering ${\cal O}$.
\end{enumerate}

\section{Column generation approach}\label{sec:CG}

Column generation is a well-known approach in mathematical optimization for
the solution of linear programming problems with a large number of
variables \cite{DesLub06}. Given a general linear programming problem
\begin{equation}
  \label{eq:P}
  \begin{array}{rl}
    \min \quad & \D \sum_{j\in{\cal V}} c_j x_j \\
    \hbox{s. to} \quad &\D\sum_{j\in{\cal V}} E_j x_j = b \\
    &\D x_j \ge 0 \quad {j\in{\cal V}},
  \end{array}
\end{equation}
where $E_j\in\mathbb{R}^r$ is the vector with the contribution of variable
$x_j$ to the $r$ constraints of the problem (we assume that $|{\cal
  V}|>r$), the simplex method optimizes
\eqref{eq:P} by finding a set of variables ${\cal B} \subset {\cal V}$ (named
set of basic variables) such that: (i) $|{\cal B}|=r$; (ii) the $r$ vectors $E_j,
j\in{\cal B}$, are linearly independent; (iii) and for any variable $j \in
{\cal N}={\cal V}\setminus {\cal B}$ (named set of nonbasic variables), we
have that the values $\mu_j= c_j - \lambda\T E_j$ (named reduced costs) are
non-negative, where $\lambda= (E_{\cal B}\T)^{-1} c_{\cal B}\in\mathbb{R}^r$ is
the set of dual variables or Lagrange's multipliers of the constraints of \eqref{eq:P}, and
$E_{\cal B}$ and $c_{\cal B}$ are respectively a matrix and vector
formed by the vectors $E_j$ and coefficients $c_j$ such that $j\in{\cal B}$.

When the number of variables $|\cal V|$ is very large, we can initially consider a 
subset $\bar{\cal V} \subseteq {\cal V}$ of variables. Problem \eqref{eq:P}
can thus be solved with the simplex method replacing ${\cal V}$ by $\bar{\cal V}$, obtaining the
sets of basic and nonbasic variables $\bar{\cal B}$ and $\bar{\cal N}$. The
simplex method guarantees that $\mu_j\ge 0$ for $j\in\bar{\cal N}$. If in
addition $\mu_j \ge0$ for $j\in {\cal V}\setminus\bar{\cal V}$ we can
certificate that the current solution is also optimal for
\eqref{eq:P}. Otherwise, there is some $j\in {\cal V}\setminus\bar{\cal V}$
with $\mu_j<0$. Column generation then solves the subproblem
\begin{equation}
  \label{eq:CGsubprob}
  \min c_j - \lambda\T E_j,\quad j\in  {\cal V}\setminus\bar{\cal V},
\end{equation}
where $\lambda$ is the vector of dual variables provided by the previous
solution of \eqref{eq:P} using $\bar{\cal V}$, and $c_j$ represents the cost
of the variable associated
to column $E_j$. The solution of \eqref{eq:CGsubprob} provides both a new
column $E_j$ and its associated reduced cost $\mu_j$. If 
the reduced cost is non-negative, we conclude that the current solution
$(x_{\bar{\cal B}},x_{\bar{\cal N}})$ is optimal. Otherwise, if the reduced
cost is negative, we add the new column $E_j$ to the set
of already generated columns (that is, $\bar{\cal V} \leftarrow \bar{\cal V}
\cup \{j\}$), and reoptimize again \eqref{eq:P}. This procedure is repeated
until \eqref{eq:CGsubprob} provides a non-negative reduced cost.

Applying the previous procedure to the $m$-invariance problem,
the column generation approach enables us to solve the continuous relaxation of
\eqref{eq:CGprob} considering only a subset $\bar{\cal C}$ of ${\cal C}^*$:
\begin{equation}
\label{eq:CGprob-partial}
  \begin{array}{rll}
 \min \quad & \D \sum_{C \in \bar{\mathcal{C}}} w_C x_C \\
 \hbox{s. to} \quad &\D\sum_{C \in \bar{\mathcal{C}}: i \in C} x_C = 1  \quad &  i\in {\mathcal P} \\
 & x_{\cal C} \in [0,1] & {\cal C} \in \bar{\mathcal{C}},
  \end{array}
\end{equation}
where the original binary constraints $x_{\cal C} \in \{0,1\}$ have been
relaxed and replaced by $x_{\cal C} \in [0,1]$.
At each iteration we test if the solution of \eqref{eq:CGprob-partial} is
optimal for the continuous relaxation of \eqref{eq:CGprob} by solving the following optimization problem for each size $\eta \in \{m,\ldots,2m-1\}$:
\begin{equation}
  \label{eq:PPproblem}
  \begin{array}{rll}
    \min & \frac{1}{2\eta} \sum_{(i,j) \in A} D_ {ij} z_{ij} - \sum_{i \in {\mathcal P}} \frac{\lambda_i} {\eta-1} y_i \\
    \hbox{s. to}& y_i = \sum_{(j,i) \in \delta_i^-} z_{ji} + \sum_{(i,j)\in \delta_i^+} z_{ij} & i \in {\mathcal P}
    \\
    & \sum_{(i,j) \in A}  z_{ij} = \eta (\eta-1)/2 \\
    & y_i - (\eta-1) z_{ij} \geq 0 &  ij \in A \\
    & z_{ij} \in \{0,1\} & ij \in A
  \end{array}
\end{equation}
where $A = \{(i,j)| i,j \in {\mathcal P}, i<j, c(i)\neq c(j)\}$, $\delta_i^+=\{(i,j) \in A\}$, $\delta_i^-=\{(j,i) \in A\}$, and $\lambda_i$ is 
the value of the dual variable with respect to
constraint for point $i$ in \eqref{eq:CGprob-partial}. The objective function
\eqref{eq:PPproblem} is the reduced cost of a new feasible cluster represented by
binary variables $z_{ij}$ (which are 1 if points $i,j$ are in the cluster,
and 0 otherwise).

Problem~\eqref{eq:PPproblem} is solved  by adapting the method described in \cite{CasGenSpa22a} to the $m$-invariance case
by simply fixing to zero $z_{ij}$ when $(i,j) \notin A$.

Within the decomposition approach of 
Section \ref{sec:decomposition}, problem \eqref{eq:CGprob-partial} is solved for each 
 subset of points ${\mathcal P}_k$, $k=1,\dots,s$, and a subset $A_k$ of $A$ is
defined accordingly. 

The previous column generation algorithm provides an optimal solution to the
continuous relaxation of \eqref{eq:CGprob}. If all the variables $x_C$ are
either 0 or 1, this solution is optimal for the integer optimization model
\eqref{eq:CGprob}. If some variables $x_C$ are fractional, some rounding
heuristic is needed to obtain a (suboptimal but in general good quality) binary
solution. In most cases, the best binary solution was obtained by solving 
\eqref{eq:CGprob-partial} with the last set of clusters $\bar{\mathcal{C}}$
computed, and replacing bounds $x_C \in [0,1]$ by binary constraints $x_C \in \{0,1\}$.

\section{Computational tests}

The column generation algorithm for $m$-invariance  introduced in this
paper has been implemented in C++. The solution of the linear optimization
problems \eqref{eq:CGprob-partial} and integer optimization subproblems
\eqref{eq:PPproblem} of the column generation algorithm were
computed with the CPLEX solver. Alternatively, subproblems
\eqref{eq:PPproblem} can also be solved, for small values of $m$, using the
implicit enumeration scheme of \cite{AloHanRocSan14}.
A parallel version of Step 4 of the decomposition approach of Section \ref{sec:decomposition} was
implemented using OpenMP.

The implementation was tested with
the ``Adult''  \cite{AdultDataset} and the ``IPUMS USA'' \cite{IPUMS} datasets.
Both datasets have been used in several previous works on
syntactic privacy for dynamic data publishing.  From the ``Adult'' dataset the
attributes  age, sex and education-num have been used as quasi-identifiers and
occupation as sensitive attribute (that is, ``occupation'' is the ``color'' attribute according
to the notation of Section \ref{sec:IPmodels}). For the ``IPUMS'' dataset we considered a data extract with columns sex, age, educ and occupation, using occupation as sensitive attribute and the rest as quasi-identifiers. We used samples of 1500 and 1000 randomly selected users for Adult and IPUMS respectively.

These two datasets were considered for their prevalent appearance in the
related literature and to reflect two scenarios depending on the number of
unique sensitive values. The Adult dataset has 13 unique sensitive values
while IPUMS has 281 unique sensitive values. We tested the performance of our
approach in comparison with the implementations of assignment and partitioning
of \cite{2007} (denoted as ``Classic''), and \cite{2017} (denoted as ``Tau''). These implementations are the two main approaches in the existing literature. 

\newcommand{\ra}[1]{\renewcommand{\arraystretch}{#1}}

\begin{table*}[ht]
    \center
    \ra{1.3}
    \caption{Computational results for the Adult dataset}
    \begin{tabular}{cc c  lcc c  ccc c  cc}
       \multicolumn{1}{c}{ }  & \multicolumn{1}{c}{ }& &\multicolumn{3}{c}{$m$-Invariance Heuristics}& & \multicolumn{3}{c}{Optimization} && \multicolumn{2}{c}{Two-Swapping} \\
      \cline{4-6} \cline{8-10} \cline{12-13} 
       Dataset & Cluster Size $m$ & &Algorithm & Time & \emph{IL} &&  s & Time & \emph{IL} && Time & \emph{IL}\\\hline
        Adult & 3 && Classic & 0.14 & 66.76&  & 40 & 0.08 & 15.60 && 41.12 & 2.36 \\
        & &   &&  &  && 20 & 0.20 & 10.49 & &37.36 & 2.17 \\
        & & &Tau & 0.02 & 39.03& & 10 & 4.02  & 7.47 & &33.62 & 2.15 \\
        & & && &  & &5 & 18.21 & 5.49 & &27.22 &2.14 \\
        & & && & & &2 & 83.00 & 3.13 & &17.73 & 1.87 \\ \cmidrule{2-13}
        & 5 && Classic & 0.07 & 81.46& & 40 & 2.66 & 29.73 && 89.49 &8.86\\
        &  &&  &  &  && 20  & 89.08 & 23.83 && 87.72 & 9.61 \\
        &  && Tau & 0.14 & 51.84& & 10  & 2872.1 & 18.34 && 67.38 & 9.06 \\
        & && & & && 5 & 5068.5 & 19.61 && 76.76 & 8.92 \\
        & && & & && 2  & 7203.2 & 21.93 && 57.28 & 8.50 \\ \cmidrule{2-13}
        & 7 && Classic & 0.37 & 88.73&   & 40  &  989.26 & 50.82 && 197.97 &22.25\\
        &  &&  &  &  & & 20  & 1357.7 & 47.68 && 178.41 &22.79  \\
        &  && Tau & 0.12 & 57.97 && 10  & 5543.0 & 46.81 && 147.25 & 22.58 \\
        & && & & & &5 & 5080.1 & 47.84 && 146.58 & 22.43 \\
        & && & & & &2 & 3929.7 &47.33 && 141.26 & 22.44 \\ 
         \hline
    \end{tabular}
    \label{table:adult}
\end{table*} 

\begin{table*}[ht]
    \center
    \ra{1.3}
    \caption{Computational results for the IPUMS dataset}
    \begin{tabular}{cc c  lcc c  ccc c  cc}
       \multicolumn{1}{c}{ }  & \multicolumn{1}{c}{ }& &\multicolumn{3}{c}{$m$-Invariance Heuristics}& & \multicolumn{3}{c}{Optimization} && \multicolumn{2}{c}{Two-Swapping} \\
      \cline{4-6} \cline{8-10} \cline{12-13} 
       Dataset & Cluster Size $m$ & &Algorithm & Time & \emph{IL} &&  s & Time & \emph{IL} && Time & \emph{IL}\\\hline
        IPUMS & 3& & Classic & 0.14 & 66.76&  & 40 & 0.06 & 11.91 && 10.57 & 0.75 \\
        & &&   &  &  && 20 & 0.09 & 6.10 && 9.17 & 0.77 \\
        & && Tau & 0.02 & 39.03 && 10 & 0.17  & 3.31 && 7.49 & 0.54 \\
        & && & &  && 5 & 0.80 & 1.57 && 5.86 & 0.55 \\
        & && & & && 2 & 4.45 & 0.74 && 2.66 & 0.49 \\ \cmidrule{2-13}
        & 5 && Classic & 0.07 & 81.46 && 40 & 0.13 & 26.22 && 23.67 &1.70\\
        &  &&  & &  && 20  & 0.86 & 13.93 && 0.86 & 1.38 \\
        &  && Tau & 0.14 & 51.84 && 10  & 3.39 & 6.94 && 19.19 & 1.21 \\
        & && & & && 5 & 48.61 & 3.63 && 14.45 & 1.19 \\
        & && & & && 2  & 181.33 & 1.37 && 7.07 & 0.86 \\ \cmidrule{2-13}
        & 7 && Classic & 0.37 & 88.73   && 40  &  910.71 & 37.48 && 44.15 & 2.28\\
        &  &&  &  &  &&  20  & 888.74 & 21.68 && 49.45 & 2.26  \\
        &  && Tau & 0.12 & 57.97 && 10  & 5105.6 & 15.62 && 37.7 & 2.36 \\
        & && & & && 5 & 5042.4 & 16.07 && 28.88 & 2.33 \\
        & && & & && 2 & 3623.7 & 9.62 && 17.95 & 1.93 \\ 
         \hline
    \end{tabular}
    \label{table:IPUMS}
\end{table*}

\subsection{Results}

Tables \ref{table:adult} and \ref{table:IPUMS} show the results obtained for,
respectively, the ``Adult'' and ``IPUMS'' datasets. Each dataset was solved for
the cluster sizes $m\in\{3,5,7\}$ and number of subsets
$s\in\{40,20,10,5,2\}$, which amounts to 15 runs of the algorithm
with each dataset. The runs were carried out on a DELL PowerEdge R7525 with
two 2.4 GHz AMD EPYC 7532 CPUs (128 total cores), under a GNU/Linux operating
system (openSuse 15.3).

The columns of the tables provide results for the different
steps of the decomposition algorithm: step 1: $m$-invariance heuristics (used
to partition the set of points in step 3); step 4: optimization 
with column generation for each subset of points; step 5: two-swapping heuristic. 
The execution of the two state-of-the-art $m$-invariance heuristics
(``Classic'' and ``Tau'') is independent of $s$, then the values of
these columns are common to all the rows with the same $m$. For each
step we provide the computational time, and the information loss (\emph{IL})
obtained (which monotonically decreases with the steps). For step 1, the
results with the ``Classic'' and ``Tau'' heuristic are given; ``Tau'' always
outperformed ``Classic'' in terms of \emph{IL}, then it was
chosen as the initial clustering to partition the set of points in $s$ subsets.
Executions of the optimization stage were multithreaded using OpenMP, the number of threads
being equal to the number of subsets $s$. For this reason, times in the table
refer to ``wall-clock'' time, instead of CPU time. We set a time limit of two
hours for each column generation process, which was only reached in one of the most
difficult instance (``Adult'', $m=5$, $s=2$).

From Tables \ref{table:adult} and \ref{table:IPUMS} it is seen that the
current state-of-the-art heuristics ``Classic'' and ``Tau'' provide very poor
solutions, with large \emph{IL} values. The combination of the  optimization
step followed by the local search two-swapping heuristic significantly
improves the quality of the $m$-invariance solution. When $m$ is small it is observed that
clusterings with \emph{IL} close to 0 can be obtained. In general, the
\emph{IL} values computed for ``IPUMS'' are much better than for the ``Adult''
dataset; this is explained by the much larger number of values of the
sensitive variable (281 vs 13). It is also seen, as expected, that the smaller $s$,
the better \emph{IL}, at the expense of a larger overall solution time.

\section{Conclusions}

The new method suggested in this paper for the $m$-invariance and $\tau$-safety NP-hard
problems (based on a column generation technique used in mathematical
optimization for the solution of large-scale problems) significantly
outperformed current state-of-the-art heuristics in terms of quality of the
solution (\emph{IL}). A potential drawback of our approach is the excessive
computational time of the optimization step when $m$ is large and $s$ is
small. Reducing the solution time of the column generation algorithm would be
part of the further work to be done.


\section*{Acknowledgments}
Jordi Castro has been supported by the MCIN/AEI/FEDER project
RTI2018-097580-B-I00.

Adrian Tobar has been supported by the Spanish Government under the project ``Enhancing Communication Protocols with Machine Learning while Protecting Sensitive Data (COMPROMISE)'' PID2020-113795RB-C31, funded by MCIN/AEI/10.13039/501100011033, and through the project ``MOBILYTICS'' (TED2021-129782B-I00), funded by MCIN/AEI/10.13039/501100011033 and the European Union ``NextGenerationEU''/PRTR. 

\bibliographystyle{IEEEtran}
\bibliography{newbib}

\section{Biography Section}

\begin{IEEEbiographynophoto}{Adrian Tobar}
received a four-year BSc degree in Mathematics (Extraordinary end-of-studies Award) from the Universitat de les Illes Balears, Palma de Mallorca, Balearic Islands, in 2019, and a MSc degree in Advanced Mathematics from the Universitat Polit\`ecnica de Catalunya in 2020. He is currently PhD student at Department of Networking Engineeering  at the Universitat Politècnica de Catalunya and assistant professor at the Department of Mathematics and Computer Science of the Universitat de les Illes Balears. 
\end{IEEEbiographynophoto}

\begin{IEEEbiographynophoto}{Jordi Castro}
received a five-year BSc-MSc degree in Computer Sciences (best mark of his class) from the Universitat
Polit\`ecnica de Catalunya, Barcelona, Catalonia, in 1991, and a PhD degree
in Computers Sciences (Operations Research) from the same university in
1995. From 1996 to 1999 he was associate professor at the Universitat Rovira i
Virgili, Tarragona, Catalonia. He is currently full professor at
the Universitat Polit\`ecnica de Catalunya. His research interests
are in algorithms for large-scale optimization problems.
\end{IEEEbiographynophoto}
\vspace{11pt}

\begin{IEEEbiographynophoto}{Claudio Gentile}
received a BSc-MSc degree in Computer Sciences (mark 110/110, with honors) from the Universit\`a di Pisa, and a Diploma from the Scuola Normale Superiore of Pisa, Italia, in 1995; he also received a PhD degree in
Operations Research from University ``La Sapienza", Roma, Italia, in 2000.
He is currently Research Director at Institute for System Analysis and Computer Science ``A. Ruberti" (IASI) of the Italian National Research  Council (CNR) where he spent all his scientific carrier. His main research interests are in Combinatorial Optimization, Polyhedral Theory, and Mixed-Integer Nonlinear Programming. 
\end{IEEEbiographynophoto}
\vspace{11pt}

\vfill

\end{document}